\documentstyle[aps,prl]{revtex}
\begin{document}
\input epsf.sty

\twocolumn[\hsize\textwidth\columnwidth\hsize\csname %
@twocolumnfalse\endcsname

\title{Crossover Phase Diagram of 2D Heisenberg Ferro- and Antiferromagnets}

\author{Alexander Sokol}
\address{Department of Physics, University of Illinois at
Urbana-Champaign, Urbana, IL 61801-3080\\
and L.D. Landau Institute, Moscow, Russia}
\author{Norbert Elstner}
\address{Service de Physique Th\'{e}orique, CEA-Saclay, 91191
Gif-sur-Yvette Cedex, France}
\author{Rajiv R.P. Singh}
\address{Department of Physics, University of California, Davis, CA
95616}

\date{cond-mat/9505148, May 29, 1995}

\maketitle

\begin{abstract}
We propose a single crossover phase diagram applicable to
2D collinear Heisenberg antiferromagnets (AFMs) and
ferromagnets (FMs), and show that the scaling regimes of AFMs and FMs
are in one-to-one correspondence. The phase diagram is split into
classical and quantum regions. Our two key results are:
(i) in the classical region,
the AFMs and FMs exhibit nearly identical behavior near their respective
ordering wavevectors, which we observe for S=1 and higher using series
expansions;
and (ii) in the quantum region, quantum critical (QC) regime
is present not only for the AFMs, but for FMs as well.
\end{abstract}

\vspace{0.05in}

\phantom{.}
]

\narrowtext
\pagebreak
\section{Introduction}

We report a study of two-dimensional, collinear,
spin-S antiferromagnets (AFMs) and ferromagnets (FMs) aimed at
better understanding the magnetic behavior of both, and in particular
the role of quantum versus classical fluctuations in these systems.
A number of quasi-$2D$ experimental systems, including spin-$\frac{1}{2}$ AFMs
La$_2$CuO$_4$ and Sr$_2$CuO$_2$Cl$_2$ \cite{spin1half,spin1half:RCvsQC} ,
spin-1 AFMs La$_2$NiO$_4$ and K$_2$NiF$_4$ \cite{spin1},
spin-$\frac{5}{2}$ AFM Rb$_2$MnF$_4$ \cite{spin5half} as well as
spin-$\frac{1}{2}$ FMs such as K$_2$CuF$_4$ \cite{dejongh} are
well described over a range of temperatures by the $2D$ Heisenberg model
\begin{equation}
H = J\sum_{\langle ij\rangle} {\bf S}_i {\bf S}_j
\label{H}
\end{equation}
on a square lattice, which we study by high temperature
series expansion methods. We expect much of our results to
apply to other lattices having collinear long-range order at $T=0$.

Let us begin by considering the relevant energy scales in the problem.
One important energy scale for both AFMs ($J>0$) and FMs ($J<0$) is
the $T=0$ spin stiffness $\rho_s$, defined as rigidity with respect to
a twist in the magnetic structure.
Quantum $1/S$ corrections
make this quantity different for AFMs and FMs with the same value of
spin; for model (\ref{H}), the order of magnitude is however the same,
$\rho_s^{\rm AFM}\sim JS^2$, $\rho_s^{\rm FM}=JS^2$
(note that the FM value is exact).

As shown by Chakravarty, Halperin, and Nelson  \cite{CHN} for
AFMs, and by Kopietz and Chakravarty \cite{KC} for FMs,
the asymptotic $T\to0$ magnetic behavior for these models obeys scaling and
depends on only two dimensionful parameters:
the energy scale $\rho_s$, and a quantity which sets the
overall length scale and can be obtained from the $q\to0$ limit of
the spin wave dispersion $\epsilon(q)$. For AFMs, $\epsilon(q)$ is
linear and one defines the $T=0$ spin wave velocity as
$c=\lim_{q\to0}\epsilon(q)/q\sim JSa$.
For FMs, $\epsilon_q$ is quadratic and one defines the $T=0$ spin wave
stiffness $\rho = \lim_{q\to 0} \epsilon(q)/q^2\sim JSa^2$.
Note that the spin wave stiffness $\rho$
has the dimension of $\mbox{energy}\times\mbox{length}^2$, while
the spin stiffness $\rho_s$ has the dimension of energy.

For the Heisenberg model, the spin wave spectrum is
well-defined for all $q$ and its upper bound, reached at the boundary
of the Brillouin zone where $q\sim 1/a$,
can be estimated as
$\Lambda \sim c/a \sim JS$ for the AFMs, or
$\Lambda \sim \rho/a^2 \sim JS$ for the FMs, i.e. not only
$\rho_s$, but also $\Lambda\sim JS$ has the same order of magnitude in
AFMs and FMs.

Whereas the $T\to0$ behavior depends on $\rho_s$ and $c$ for AFMs,
or $\rho_s$ and $\rho$ for FMs,
the $T\to\infty$ behavior (the Curie-Weiss law) depends on
$\rho_{\rm CL}=JS(S+1)$ and the lattice constant $a$.
In what follows, we show
that this simple observation is part of a general picture where
the low temperature behavior, which may contain several scaling regimes,
at $T\sim \Lambda$ crosses over to
the high-temperature behavior,
which may also contain several scaling regimes.
A similar crossover in 3D occurs {\em below} the long
range ordering temperature (which is zero in 2D), and was studied by
Vaks, Larkin, and Pikin \cite{Vaks:Larkin:Pikin}.

In model (\ref{H}), the ratio $\Lambda/\rho_s\sim 1/S$ can be
made arbitrarily small by increasing $S$, but it cannot
be made arbitrarily large as $S\ge \frac{1}{2}$. Nevertheless, the case
$\Lambda/\rho_s\gg 1$ is of great interest.
First, there exist models where
$\rho_s$ goes to zero while $\Lambda$ does not,
in which case $\Lambda/\rho_s\gg 1$ applies rigorously;
an example of such a system is the two-layer
Heisenberg model \cite{Sandvik:Scalapino,SCS} near the critical point
where the N\'eel AFM long-range order vanishes.
Second, for the $S=1/2$ AFM model (\ref{H}),
the ratio $\Lambda/\rho_s \sim 10\gg1$ is quite large; furthermore,
there exists evidence
\cite{CSY,SGS,Sandvik:Scalapino,EGSS,SCS}, first pointed out by
Chubukov and Sachdev, that at intermediate temperatures this model
is indeed in the quantum critical limit $\Lambda/\rho_s\gg 1$.

\newpage
\phantom{.}
\vfill
\begin{figure}
\centerline{\epsfxsize=3.0in\epsfbox{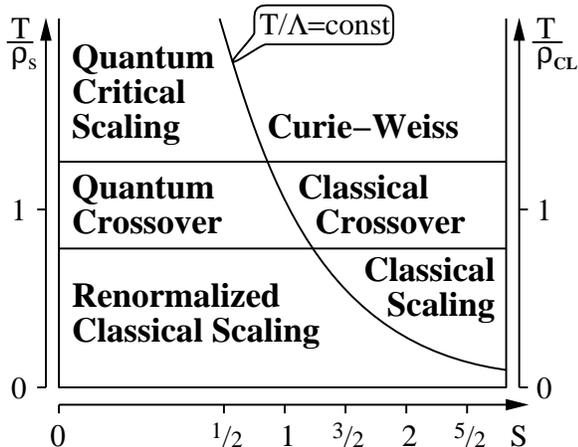}}
\caption{A phase diagram of spin-S quantum antiferromagnets (AFMs)
and ferromagnets (FMs); see Table I for the corresponding scaling and
crossover expressions for the correlation length.
All regime boundaries are gradual crossovers rather
than phase transitions, and their positions can only be defined within
numerical factors of order unity.
The behavior in the $T/\Lambda\gg 1$ region
is in the universality class of 2D classical magnets,
where magnetic properties of AFMs and FMs for the same value of $S$ are
the same near their respective ordering wavevectors, and depend on
$T/\rho_{\rm CL}$, where $\rho_{\rm CL}=JS(S+1)$. The classical
behavior includes
Curie-Weiss regime ($T\gg \max(\rho_s,\Lambda)$) and
classical scaling regime ($\Lambda\ll T\ll \rho_{\rm CL}$), which
are separated by a fairly wide {\em classical crossover}
regime for $\Lambda \ll T\sim \rho_{\rm CL}$,
where scaling does not hold, but
nevertheless $\xi_{\rm AFM}\approx \xi_{\rm FM}\approx a
\psi_{\rm CL}(T/JS(S+1))$.
In the region $T/\Lambda\ll 1$
quantum effects are important and therefore AFMs and FMs behave
differently. For the AFMs, this region is in the universality class of
the QNL$\sigma$ model. Here these scaling regimes are found
for both AFMs and FMs:
renormalized classical (RC) regime for $T\ll \min(\rho_s,\Lambda)$;
quantum critical (QC) regime for $\rho_s\ll T\ll \Lambda$; and a wide
quantum
crossover regime for $T\sim \rho_s\ll \Lambda$, where the correlation
length remains a universal function (different for AFMs and FMs)
of the thermal de Broglie
wavelength for spin waves ($\lambda_T=c/T$ or
$\lambda_T=(\rho/T)^{1/2}$) and the ratio $T/\rho_s$.
}
\label{fig:pd}
\end{figure}

\vfill
\phantom{.}
\newpage
\phantom{.}
\vfill

\begin{center}
\begin{tabular}{|l|l|}
\hline
Regime & Correlation Length \\
\hline
\hline
Curie-Weiss                  & $\xi_{\rm AFM} = \xi_{\rm FM}\alt a$ \\
$T \gg \max(\rho_{\rm CL},
 \Lambda)$                   & Properties depend on $T/\rho_{\rm CL}$ \\
\hline
Classical Crossover          & $\xi_{\rm AFM} = \xi_{\rm FM}$ \\
$T\sim \rho_{\rm CL}\gg
 \Lambda$                    & $\ \ = a \psi_{\rm CL}(T/\rho_{\rm CL})$ \\
\hline
Classical                    & $\xi_{\rm AFM} = \xi_{\rm FM}
                               \sim a (T/\rho_{\rm CL}) $ \\
$\rho_{\rm CL}\gg T \gg
\Lambda$                     & $\ \ \times \exp(2\pi\rho_{\rm CL}/T)$ \\
\hline \hline
Quantum Critical             & $\xi_{\rm AFM} \sim c/T$, \\
$\rho_s\ll T\ll \Lambda$     & $\xi_{\rm FM} \sim
                               \left(\rho/T\log(T/\rho_s)\right)^{1/2} $ \\
\hline
Quantum Crossover            & $\xi_{\rm AFM} = (c/T) \,
                               \phi_{\rm AFM}(T/\rho_s)$, \\
$T\sim \rho_s \ll \Lambda$   & $\xi_{\rm FM} = (\rho/T)^{1/2}
                               \phi_{\rm FM}(T/\rho_s)$ \\
\hline
Renormalized Classical       & $\xi_{\rm AFM} \sim (c/\rho_s) \,
                               \exp(2\pi\rho_s/T)$, \\
$T \ll \min(\rho_s,\Lambda)$ & $\xi_{\rm FM} \sim (\rho T/\rho_s)^{1/2}
                               \exp(2\pi\rho_s/T)$ \\
\hline
\end{tabular}
\end{center}
\noindent {\small
{\sc Table I}. Correlation length in the scaling and crossover regimes
shown in the phase diagram of Fig.\protect\ref{fig:pd}, after
Refs.\cite{Luscher:Weisz,CHN,KC,CSY}, and this work. The sign
$\sim$ indicates the presence of a numerical prefactor. The functions
$\phi_{\rm AFM}$ and $\phi_{\rm FM}$ are universal; the function
$\psi_{\rm CL}$ is not, for instance,
it explicitly depends on the ratio of the next-nearest and
nearest-neighbor exchange couplings.
}
\bigskip

\vfill
\phantom{.}
\newpage

\section{Curie-Weiss regime}

At high temperatures $T\gg \max(\rho_s,\Lambda)$,
the spins are weakly correlated and one can keep only a few leading terms
in the high temperature series
expansion. For model (\ref{H}),
this temperature range corresponds to $T\gg \max(JS,JS^2)$.
The corresponding
mean-field theory yields the familiar Curie-Weiss law.

\section{Renormalized Classical Regime}

In the opposite limit of $T\ll \min(\rho_s,\Lambda)$,
the behavior is of the
renormalized classical (RC) type. This regime was studied in detail in
Refs.\cite{CHN,KC,CSY},
where magnetic properties were calculated by mapping
the system onto the classical nonlinear $\sigma$-model and setting the
momentum-integration cutoff to be
proportional to the thermal de Broglie wave vector $q_T$, defined such
that $\epsilon(q_T)\sim T$.
The predicted $\xi(T)$ has the same form for AFMs ($J>0$) and FMs ($J<0$),
when expressed in terms of their respective thermal
de Broglie wavelengths
$\lambda_T=1/q_T$:
\begin{equation}
\xi_{\rm RC} =\frac{e}{8} \lambda_T \frac{T}{2\pi\rho_s}
\exp\left(\frac{2\pi\rho_s}{T}\right),
\label{xi:rc}
\end{equation}
(here the exact value of the prefactor is obtained from the correlation length
of the classical $O(3)$ nonlinear-$\sigma$ model in the minimal subtraction
scheme \cite{HN,JLG}).
However, not only the values of $\rho_s$ differ for FMs and AFMs, but
also $\lambda_T$ has qualitatively different temperature dependences:
\begin{equation}
\lambda_T = \left\{
\begin{array}{ll}
c/T & \ \ \mbox{for AFMs} \\
\sqrt{\rho/T} & \ \ \mbox{for FMs}
\end{array}
\right. .
\label{lambdaT}
\end{equation}

An important question is upto what
temperatures should this RC expression hold?
We recall that
when $T$ is larger than the upper bound
of the $T=0$ spin wave spectrum, $\Lambda$, there can be no spin waves with
wavelengths that are comparable to the thermal de Broglie wavelength.
Since such spin waves are important for RC theory, one would expect it
to be applicable only for
$T\ll \min(\Lambda,\rho_s)$, which for model (\ref{H}) translates into
$T\sim \min(JS,JS^2)$ (here, we ignore numerical factors of order unity).
At higher temperatures new physics must arise.

The Curie-Weiss regime occurs for
$T\gg \max(\rho_s,\Lambda)$ and the RC regime for
$T\ll \min(\rho_s,\Lambda)$. These two regimes exist for all $S$.
We now turn to the discussion of the
regimes where the temperature is larger than one of the
scales $\rho_s$ or $\Lambda$,
but smaller than the other.

\begin{figure}
\centerline{\epsfxsize=3.0in\epsfbox{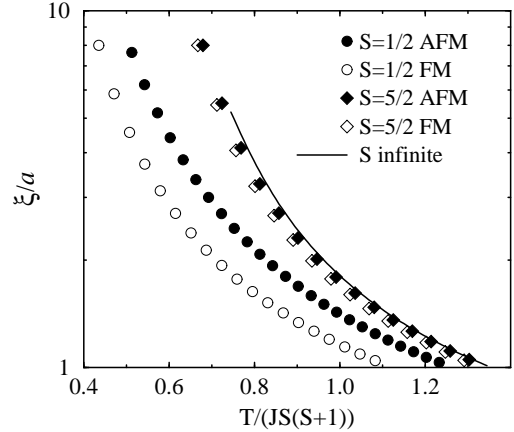}}
\caption{A semi-log plot of $\xi/a$ vs. $T/(JS(S+1))$
for AFMs (solid symbols) and FMs (open
symbols) for $S=1/2$ (circles) and $S=5/2$ (diamonds).
We also plot the data from Ref.\protect\cite{Luscher:Weisz} for the classical
$S\to\infty$ model (solid line).
For $S=1/2$, AFM and FM correlation lengths are different and
neither agrees with the $S\to\infty$ limit.
For $S=5/2$, AFM and FM
correlation lengths agree {\em both} with each other and
with the $S\to\infty$ model.
The data for $S=5/2$ is plotted as representative for
large spin models; for all studied
spins $S>1$, the behavior is similar although the agreement between
$\xi_{\rm AFM}$ and $\xi_{\rm FM}$ for $S=3/2$ and $S=2$
is not as striking as for $S=5/2$.
This data collapse provides evidence that any quantum effects, which are
expected to be different for AFMs and FMs, become unimportant already
for fairly small spins $S>1$,
in agreement with the conjecture due to Ref.\protect\cite{ESSGB}.
}
\label{fig:xi}
\end{figure}

\section{Classical and Classical Crossover Regimes}

In our earlier work in collaboration
with Greven and Birgeneau \cite{ESSGB}, we proposed a scaling
crossover scenario to describe substantial deviations from RC behavior
observed in the neutron scattering experiments for AFMs
with spin-one and larger. This scenario calls for a crossover to the
behavior of the $S\to\infty$ classical system when
spin wave energies for all wavevectors in the Brillouin zone
become smaller than the
temperature, i.e. for $T\agt \Lambda$.
The correlation length at this RC to classical boundary, obtained from
Eqs.(\ref{xi:rc}),
is exponentially large for large spin,
\begin{equation}
\left. \frac{\xi}{a}\right|_{T\sim\Lambda}
\sim \frac{\exp(S)}{S}\gg 1 \ \ \mbox{for $S\gg 1$}.
\end{equation}
In \cite{ESSGB}, the arguments in favor of RC to classical crossover
were based on data collapse for large
spin when plotted versus $T/(JS(S+1))$. The studied values of spin were
small enough ($S=1/2$ to $S=5/2$) to make the results sensitive to the
choice of the renormalized temperature as $T/(JS(S+1))$ or
$T/JS^2$.

Here, we resolve this arbitrariness by studying antiferro- and
ferromagnets simultaneously. AFMs and FMs have the same
classical limit, but the quantum effects differ and therefore the
difference between AFMs and FMs can be used as a probe for the
strength of quantum effects. It turns out that for any $S>1$,
in the temperature range of interest ($\xi\sim 10^1$),
the correlation length
calculated by high temperature expansions is nearly the same
for AFMs and FMs with the same value of $S$ (Fig.\ref{fig:xi}).
Therefore, it must be determined
by classical rather than quantum physics.
We emphasize that such an argument does not rely on any particular way
of collapsing the data,
because it is drawn from comparisons of AFM
and FM models for the same spin.
On the other hand, for $S=1/2$ the difference between $\xi_{\rm AFM}$
and $\xi_{\rm FM}$ is always large and temperature-dependent, which indicates
the importance of quantum physics for this value of $S$. Spin-one appears to
be a borderline case.

Having established that in the temperature range
$T\gg\Lambda$ finite-spin models
behave similarly to the $S\to\infty$ classical magnet,
we now comment on the behavior of the latter. The
asymptotic low-temperature behavior of the classical 2D Heisenberg
antiferromagnet is given by \cite{JLG}:
\begin{equation}
\frac{\xi_{\rm CL}}{a} = \frac{e}{8}\,
\frac{e^{-\pi/2}}{\sqrt{32}}\,
\frac{T}{2\pi \rho_{\rm CL}}\,
 \exp\left(\frac{2\pi\rho_{\rm CL}}{T}\right),
\label{classical}
\end{equation}
where $\rho_{\rm CL}=JS(S+1)$.
It turns out that because of the very small
numerical value of the prefactor ($\sim 0.01$),
this formula becomes accurate only for very low temperatures,
$T/\rho_{\rm CL} \alt 0.6$ ($\xi \agt 100$).
We label the intermediate temperature range where neither
Curie-Weiss nor classical scaling behavior apply the
{\em classical crossover} regime. In this regime,
neither large nor small-temperature asymptotic expressions describe
the correlations accurately, nevertheless, all properties of the model
are nearly the same for AFMs and FMs, and depend on $T$ through
$T/\rho_{\rm CL}$ only.

\section{Quantum Critical and Quantum Crossover Regimes}

We call {\em quantum critical}
that behavior which can formally be obtained for $\rho_s\ll T\ll \Lambda$.
For the antiferromagnets, this regime is the asymptotic {\em quantum
critical} (QC) regime \cite{CHN,CSY}.
The correlation length $\xi$ is proportional to
$\lambda_T=c/T$ with corrections that depend
universally on $T/\rho_s$, and to leading order the dominant
frequency scale for spin fluctuations is $\bar{\omega}\sim T$.

For the FMs the spin wave spectrum is quadratic, which causes
infrared ($q\to 0$) log divergences for
$\rho_s\ll T\ll \Lambda$.
Such divergences (not present for AFMs) are cutoff by $\rho_s$, and
lead to the following multiplicative correction to the correlation length:
\begin{equation}
\xi_{\rm FM} \sim \left(\rho/T\right)^{1/2} \log^{-1/2}
\left(T/\rho_s\right),
\label{QC:FM}
\end{equation}
in either the sigma model or the Schwinger boson formalism.

Note that such a behavior cannot be the true quantum critical
behavior because the limit
$\rho_s\to0$ leads to singular $\xi$ at
finite $T$. Therefore, Eq.(\ref{QC:FM}) must fail as $\rho_s\to 0$,
and the $T=0$ quantum ferromagnet-paramagnet phase transition is likely
to belong to a different universality class \cite{Sachdev:conserved}.
Nevertheless, lacking a better
name and for compatibility with the AFM nomenclature,
we call this behavior a {\em
ferromagnetic quantum critical} regime.

The applicability of the QC description to model (\ref{H}) for $S=1/2$
AFM is widely discussed in the literature
\cite{spin1half:RCvsQC,CSY,SGS,Sandvik:Scalapino,EGSS}. A similar
analysis for the ferromagnets requires detailed calculations of their
properties in the QC regime, and is beyond the framework of this paper.

\section{Conclusion}

In this paper, we present a complete phase diagram of spin-S
antiferromagnets (AFMs) and ferromagnets (FMs).
We show that comparing AFMs and FMs allows one to elucidate
crossover effects from the quantum regimes (including the renormalized
classical regime where, despite its name, quantum fluctuations are
essential) to the purely classical regimes, where AFMs and FMs exhibit
identical behavior near their respective ordering wavevectors. For a
detailed description of the phase diagram and the proposed new
regimes, we refer the reader to Fig.\ref{fig:pd} and Table 1.

For the antiferromagnets,
both our series expansion calculations and the
neutron data \cite{spin1half:RCvsQC,ESSGB} suggest that
the regime boundaries in Fig.\ref{fig:pd} are positioned
such that the universal $T\ll \Lambda$
scaling theory, free of any lattice corrections,
is not obeyed for spin-one and larger at any numerically or
experimentally accessible values of the correlation length.
The data also highlights the disagreement in the magnitude of
spin correlations even for $S=1/2$ at all accessible temperatures with
the existing RC predictions, whereas
$\xi(T)$ was found \cite{spin1half:RCvsQC} to agree
remarkably well with the RC theory.

Our results suggest that an accurate
analytical theory for $S\geq1$
in the experimentally relevant temperature range
can be constructed by taking into account
leading quantum corrections about the $S=\infty$ limit on a lattice.
The important step here would be to obtain an analytic approximation
which is valid in the {\it classical crossover regime}.
On the other hand,
any theory based on a purely continuum
description, such as the QNL$\sigma$ model without
lattice corrections, is clearly
inadequate for any spin larger than $S=1/2$.

We are grateful to R.J. Birgeneau, S. Chakravarty,
A.V. Chubukov, M. Greven, Th. Jolic\oe ur, D. Pines,
S. Sachdev, and S. Sondhi for many useful discussions.
This work is supported by the NSF Grant No. DMR 93-18537.
One of us (A.S.) is supported by an Alfred P. Sloan Research Fellowship.



\begin{references}

\bibitem{spin1half}
B. Keimer {\em et al.}, Phys. Rev. B  {\bf 46}, 14034 (1992);
R.J. Birgeneau {\em et al.} (unpublished)

\bibitem{spin1half:RCvsQC}
M. Greven {\em et al.}, Phys. Rev. Lett.  {\bf 72}, 1096 (1994),
and Z. Phys. B {\bf 96}, 465 (1995)

\bibitem{spin1}
K. Nakajima {\em et al.}, Z. Phys. B {\bf 96}, 479 (1995);
R.J. Birgeneau, Phys. Rev. B  {\bf 41}, 2514 (1990),
and references therein.

\bibitem{spin5half}
R.A. Cowley {\em et al.}, Phys. Rev. B {\bf 15}, 4292 (1977).

\bibitem{dejongh} L. J. Dejongh and A. R. Midema, Advances in Physics,
{\bf 23}, 1 (1994).

\bibitem{CHN}
S. Chakravarty, B.I. Halperin, and D.R. Nelson,
Phys. Rev. B {\bf 39}, 2344 (1989). The exact value of the prefactor
in the expression for $\xi(T)$ has been calculated in Ref.\cite{HN}.

\bibitem{KC}
P. Kopietz and S. Chakravarty,
Phys. Rev. B {\bf 40}, 4858 (1989).

\bibitem{Vaks:Larkin:Pikin}
V.G. Vaks, A.I. Larkin, and S.A. Pikin,
Sov. Phys. JETP {\bf 26}, 188 (1968).

\bibitem{Sandvik:Scalapino}
A.W. Sandvik and D.J. Scalapino, Phys. Rev. Lett. {\bf 72}, 2777
(1993); A.W. Sandvik and A.V. Chubukov, unpublished.
A.J. Millis and H. Monien [Phys. Rev. Lett. {\bf 70}, 2810 (1994)]
proposed to use the two-layer Heisenberg model to describe
bilayer interaction in YBCO materials.

\bibitem{SCS}
A.W. Sandvik, A.V. Chubukov, and S. Sachdev, preprint (1995).
Available as cond-mat/9502012.

\bibitem{CSY}
A.V. Chubukov, S. Sachdev, and J. Ye, Phys. Rev. B {\bf 49},
11919 (1994); A.V. Chubukov and S. Sachdev, Phys. Rev. Lett. {\bf 71},
169 (1993).

\bibitem{SGS}
A. Sokol, R.L. Glenister, and R.R.P. Singh, Phys. Rev. Lett.,
{\bf 72}, 1549 (1994).

\bibitem{EGSS}
N. Elstner, R. L. Glenister, R. R. P. Singh, and A. Sokol,
Phys. Rev. B {\bf 51}, 8984 (1995).

\bibitem{Luscher:Weisz}
M. L\"uscher and P. Weisz, Nucl. Phys. B {\bf 300}, 325 (1988);
S.H. Shenker and J. Tobochnik, Phys. Rev. B {\bf 22},
4462 (1980);
D.N. Lambeth and H.E. Stanley, Phys. Rev. B {\bf 12}, 5302 (1975).

\bibitem{HN}
P. Hasenfratz M. Maggiore and F. Niedermayer, Phys. Lett. B {\bf 245},
522 (1990);
P. Hasenfratz and F. Niedermayer, Phys. Lett. B {\bf 268}, 231 (1991).

\bibitem{JLG}
Th. Jolic\oe ur and J.C. LeGuillou, Mod. Phys. Lett. B {\bf 5}, 593 (1991).

\bibitem{ESSGB}
N. Elstner, A. Sokol, R.R.P. Singh, M. Greven, R.J. Birgeneau,
preprint (1995).

\bibitem{Sachdev:conserved}
S. Sachdev, Z. Phys. B {\bf 94}, 469 (1994)

\end{references}
\end{document}